# Reusable Services from the neuGRID Project for Grid-Based Health Applications

Ashiq ANJUM[1,a], Peter BLOODSWORTH[a], Irfan HABIB[a], Tom LANSDALE[a], Richard McCLATCHEY[a], Yasir MEHMOOD[a] & The neuGRID CONSORTIUM[b]
[a]*Centre for Complex Cooperative Systems, BIT, UWE Bristol, UK*
[b]*See acknowledgements section for further details regarding neuGRID partners.*

**Abstract.** By abstracting Grid middleware specific considerations from clinical research applications, re-usable services should be developed that will provide generic functionality aimed specifically at medical applications. In the scope of the neuGRID project, generic services are being designed and developed which will be applied to satisfy the requirements of neuroscientists. These services will bring together sources of data and computing elements into a single view as far as applications are concerned, making it possible to cope with centralised, distributed or hybrid data and provide native support for common medical file formats. Services will include querying, provenance, portal, anonymization and pipeline services together with a 'glueing' service for connection to Grid services. Thus lower-level services will hide the peculiarities of any specific Grid technology from upper layers, provide application independence and will enable the selection of 'fit-for-purpose' infrastructures. This paper outlines the design strategy being followed in neuGRID using the glueing and pipeline services as examples.

## 1. Introduction

The medical domain is increasingly becoming more compute and data intensive and this has led to the adoption of distributed computing infrastructures including Grid-based architectures. In order to facilitate analysis, querying and collaboration over such infrastructures, numerous projects in the domain have built community specific high-level distributed services. These include pipeline management, provenance, image handling and anonymization services [1, 2]. Such services will help medical users in sharing data and knowledge and will enrich medical decision support systems. Most of these services are DICOM-aware Grid services designed and developed for a particular community of medical users. The services built by one community, however, cannot be readily re-used in other medical domains due to architecture, interface or platform dependencies.

The aim of the neuGRID project is to provide a user-friendly grid-based e-infrastructure, which will enable the European neuroscience community to carry out research that is necessary for the study of degenerative brain diseases. Following t he recommendations of the SHARE project's HealthGrid Roadmap [3], one of its principal goals is to provide a set of generic services, specified in consultation with its clinician user communities and thereby tailored for medical informatics that can be reusable both across Grid-based neurological data and for wider medical analyses. The services will include, but will not be limited to, query, pipeline, provenance, glueing and abstraction, anonymization and portal services and will glue a wide range of user applications to the available Grid platforms, thereby creating a foundation of cross-community and cross-platform services. For the purposes of this paper emphasis will be placed on the pipeline specification and glueing services as examples of the neuGRID services.

---

[1] Corresponding Author: Ashiq Anjum, Centre for Complex Cooperative Systems, Bristol Institute of Technology, University of the West of England, Frenchay Campus, Coldharbour Lane, Bristol BS16 1QY, Email:ashiq.anjum@cern.ch.

Numerous other related efforts exist, however none of these state-of-the-art projects satisfy all the neuGRID user requirements. Neurolog [4] is one such project, which provides a service based workflow system for neuro-imaging research. The workflow system is based on Taverna [5]. A service based workflow environment requires all tasks of a workflow to be exposed as services. NeuGRID pipelines mostly comprise neuro-imaging algorithms, which are executable tasks. Transforming these tasks into services would be expensive and at the same time would constrain scheduling in the Grid infrastructure. Similarly, caGrid [6] is a data oriented service Grid whose primary application is federating data across cancer hospitals, providing services for querying, and analyzing them using a WSRF based SOA infrastructure. In caGrid, only those sites having analysis services can carry out computation on behalf of other users in the Grid, whereas in the neuGRID architecture, any site that has all the pipeline actors installed can be used in the enactment of a pipeline. This leads to the requirement for a more scalable system since load is effectively shared amongst sites.

Triana [7] allows the authoring of both task and service based workflows and uses a component analogous to the neuGRID Glueing Service. For enacting service-based workflows Triana uses a Grid Application Prototype interface, while for task-based workflow enactment the Gridlab GAT API [8] is used. Both interfaces allow new bindings to be plugged in when they become available. To provide reusability the neuGRID Glueing service must follow the OGF standardized SAGA which originates from the GAT specification. The Gridlab GAT does not support most of the Grid infrastructures like the gLite middleware for neuGrid. Moreover, Triana does not support global workflow planning before enacting a Grid workflow.

In this paper, Section 2 highlights the salient features of the neuGRID design philosophy. Section 3 describes some requirements behind the glueing and workflow services as an example of the user-led service specification. Section 4 briefly discusses the role of glueing service in achieving Grid middleware independence and in section 5 the neuGRID approach for the construction, gridification and enactment of the pipelines is discussed. The conclusion and possible future directions are listed in section 6.

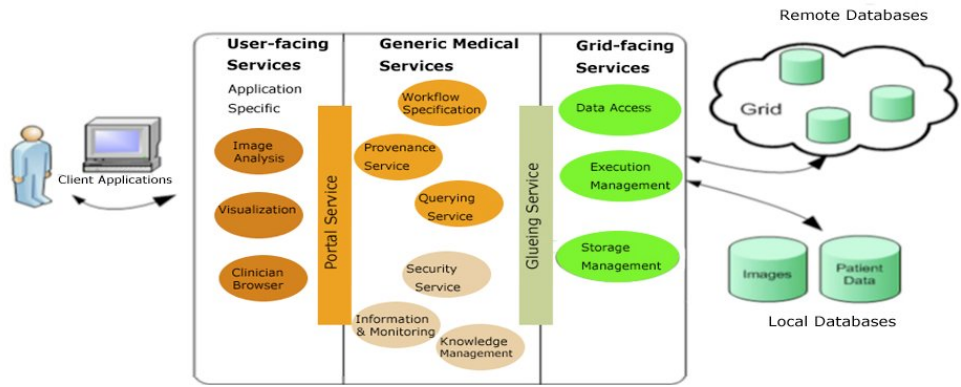

**Figure 1.** The neuGRID Layered Service Architecture

## 2. Service Design Philosophy

The neuGRID designers have adopted a layered approach by isolating user facing functionality from the backend Grid infrastructure, to deliver sets of services that address the requirements of its user community and, at the same time, maximizing the possibility of service reuse. Figure 1 shows diagrammatically this *separation of concerns* which originated from the requirements and is the fundamental principle behind the implementation of neuGRID services. In this section, we detail the philosophy which is driving the design of the distributed services in neuGRID.

A widely used and increasingly accepted standard form of a generalized distributed architecture which can support a wide range of applications and platforms is a service-oriented architecture (SOA). Extensive literature exists which shows how the main characteristics of an SOA, namely loose coupling between services, the abstraction from technological aspects, their extensibility as well as reusability of services, result in extensible architectures for distributed systems [9]. The neuGRID medical services layer is planned to be implemented using the SOA paradigm in order to leverage the advantages mentioned above.

In complex distributed systems such as neuGRID, services often need scalability for handling increased loads. To scale the neuGRID services three possible approaches have been considered, which are:

- Creating well-formed scalable services, which are stateless at the same time.
- Enabling service redundancy.
- Implementing data caching techniques.

The services being created in neuGRID are middleware agnostic and will shield the heterogeneity of distributed resources through a common abstraction layer. This will enable the services to access Grid- and non-Grid resources alike without getting locked in to a particular Grid middleware. Grid is evolving and new features are being added to Grid middleware every now and then. The service interfaces need changes with every new release of a middleware to cope with the middleware evolution. The middleware agnostic services will not have to undergo these changes since an abstraction layer will shield the services when the underlying middleware evolves.

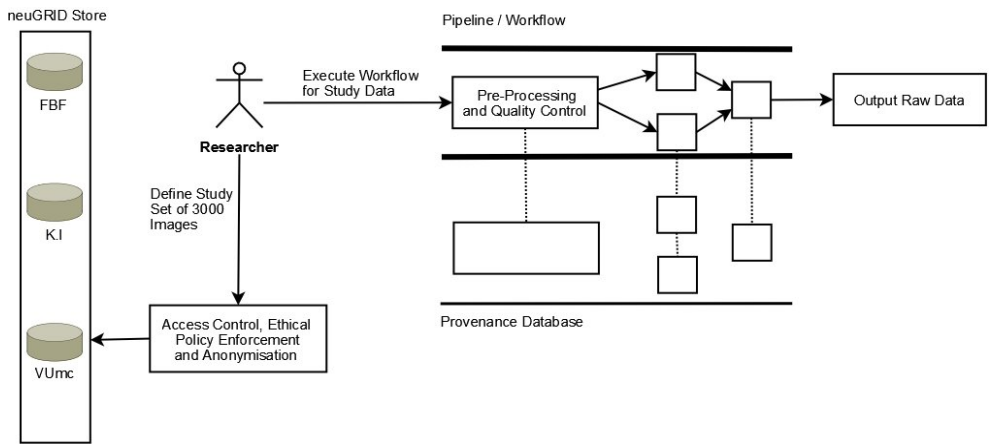

**Figure 2.** Use Case: the execution of image analysis pipelines

## 3. Requirements for Pipeline and Glueing Services

The development approach being followed in neuGRID has a high degree of end-user involvement in order to guarantee a good match between community need and technology provision. During the first year of the project a detailed set of user requirements has been gathered via a number of elicitation sessions with clinicians from partner sites. Following an extensive analysis of these requirements a set of use-cases has been established to drive the design process. Full details of the requirements analysis can be requested from the authors.

In order to illustrate how the requirements that have been elicited from the neuGRID user community are being realized by a set of generic, distributed services consider the use case presented in figure 2. In this use case the clinical researcher executes one or more pipelines of algorithms on a defined study set of images in order to satisfy a particular medical imaging analysis and to produce a set of processed output data. In order to exercise this use case the researcher may have specified that pipeline using some graphical editing tool and would have defined the content and verified the consistency and homogeneity of the study set using other potential neuGRID services (e.g. Portal Service, Querying Service, Pipeline Specification Service, Anonymization Service), but these are not considered hereinafter.

As the pipeline is processed, one or more of its elements will result in output data being produced (either transitory or persistent in nature) and information regarding the status of the pipeline execution being captured. Such descriptive data is often referred to as provenance data and a Provenance Service is responsible for storing, managing and retrieving this data. (Of course other provenance data will be collected over time, such as the definition/ownership of study sets, the history of various output data sets etc.) Thus the execution of pipelines and the collection and management of provenance data are closely linked. Furthermore the execution of the pipeline is linked to the infrastructure for the management of data and compute resources. In neuGRID we have specified a Glueing service, which facilitates the linking (and importantly the isolation) of services such as the Pipeline Execution and Provenance Services to/from the underlying enabling infrastructure.

## 4. The Glueing Service

The neuGRID Glueing Service supports an architecture, which hides the heterogeneity of distributed resources and enables generic services to access resources without getting into the details of underlying Grid middleware or architecture. The Glueing Service aims to provide:

- A standard way of accessing Grid services without tying services and applications to a particular Grid middleware.
- A solution which extends and enhances the reusability of already developed services across domains and applications.
- A service-based approach to shield users and applications from writing complex Grid specific functionality. The user interacts with a minimal set of Grid specific APIs and the rest of the functionalities are managed by the service.

The Glueing service exposes the SAGA [10] (Simple API for Grid Applications) APIs through a web service interface to meet project aims. SAGA is an open standard that is defined and maintained by the Open Grid Forum (OGF), which describes a high-level interface for simple programming of Grid applications.

The Glueing service design provides a generic framework for accessing resources over the Grid. The heterogeneity of distributed resources and details of grid middleware architectures are transparent from users. This service enables the running of a job on a particular Grid middleware by the runtime loading of a middleware adaptor. Middleware adaptors pass jobs to the middleware as its clients and are responsible for low-level communication. This mechanism shields the low-level middleware details from the user thus encouraging its use with little domain knowledge.

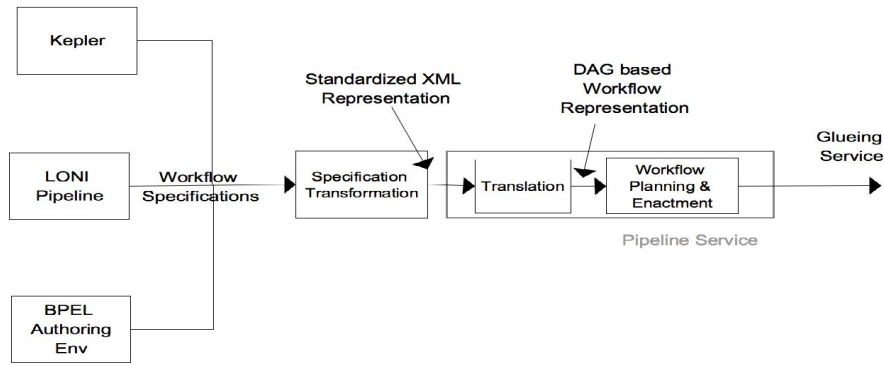

**Figure 3.** Pipeline Service Architecture

## 5. The Pipeline Specification Service

In neuGRID, the Pipeline Specification Service includes the following:

- Enable the authoring of pipelines in a user-friendly environment.
- Plan and parallelize the abstract user defined pipeline for an optimal distribution and throughput.
- Submit the pipeline to a Glueing service for its ultimate enactment and execution on a Grid and finally
- View results of the execution as well as intermediary provenance data.

As part of the Pipeline Service design specification, state-of-the-art projects were reviewed and evaluated and the architecture that most suited the neuGRID requirements was selected. The selected architecture (depicted in figure 3) has numerous features, which make it suitable for neuGRID. The architecture promotes a separation of interests, where the authoring and planning environment is completely decoupled from the gridification and enactment engine (through the Glueing Service). Numerous authoring environments can be integrated which include LONI [11], Kepler [12] or a web-based authoring interface. At the Grid end, the design integrates seamlessly with other neuGRID services, including the Glueing Service.

## 6. Conclusions and Future Work

The Distributed Service layer of the neuGRID project aims to supply a group of services to the users of the project and for reuse across other medical domains. Consequently, services have been designed that are neither middleware nor application specific to address the requirements that have been elicited from the project's user community. This paper has discussed the major components and requirements for two of these services, namely the glueing and pipeline services and has briefly highlighted the functionality that these services should address. The *generic* services will glue a wide-range of user applications to the available Grid platforms thereby creating a foundation of cross-community and cross-platform services and will enable the medical community to work cohesively whilst maintaining their independence. The services have been designed by following a set of design principles whose features include:

- Services are designed and built upon a standards based service-oriented architecture.
- Services will be exposed through interoperable interfaces.
- Services will not be tied to a particular technology and will be Grid middleware agnostic.
- Services will be made as scalable as is architecturally possible.

Future work includes the implementation of the services that have been identified in the service layer and their composition into different biomedical applications. This also includes measures to demonstrate the capacity and capability of the services on different Grid middleware. Currently neuGRID is developing a proof-of-concept prototype for demonstration to the Grid community (at the EGEE and HealthGrid user conferences). In addition the neuGRID project is actively pursuing collaboration both with other projects in neuroscience (including in Europe Neurolog, in Canada CBRAIN and in the US BIRN / LONI) and with other clinical researchers in other medical domains (for example cardiology, paediatrics and cancer studies) who may wish to reuse the neuGRID infrastructure and services for their analyses. Furthermore it continues to contribute actively to the Open Grid Forum activities towards promoting reuse and interoperability through use of service oriented architectures.

**Acknowledgements:**

The authors acknowledge the financial support of the Framework Programme 7 of the EC through the Grant Agreement number 211714. In addition they thank the partners in neuGRID for their contributions to this paper from Fatebenefratelli (Brescia, Italy), UWE (Bristol, UK), Maat Gknowledge (Archamps, France), VUmc (Amsterdam, Netherlands), HealthGrid (Clermont, France), Prodema (Bronschhofen, Switzerland), CFconsulting (Milan, Italy) and the Karolinska Institute (Stockholm, Sweden).

**References**

[1] F. Estrella et al, "A Service-Based Approach for Managing Mammography Data", The 11[th] World Congress on Medical Informatics (MedInfo'04) San Francisco, CA, USA. September 2004.
[2] DICOM Digital Imaging and Communications in Medicine, http://medical.nema.org.
[3] See http://roadmap.healthgrid.org/ from the SHARE project web-site http://eu-share.org/
[4] J. Montagnat et al, "NeuroLOG: a community-driven middleware design". HealthGrid'08, IOS Press, Chicago, June 2008.
[5] T. Oinn, et al, "Taverna: Lessons in creating a workflow environment for the life sciences", Concurrency and Computation: Practice and Experience, 18, 2006, pp. 1067-1100.
[6] S. Oster et al, "caGrid 1.0: An Enterprise Grid Infrastructure for Biomedical Research", Journal of the American Medical Informatics Association, March 1, 2008; 15(2): 138 – 149
[7] I. Taylor et al, "The Triana Workflow Environment: Architecture and Applications", Workflows for e-Science, pages 320-339. Springer, New York, Secaucus, NJ, USA, 2007.
[8] G. Allen et al, "GridLab: Enabling Applications on the Grid", In proceedings of Grid (GRID 2002), Springer Berlin, Heidelberg, 2002, pp.39-45.
[9] M. P. Papazoglou & W,.v. Heuvel, (2007). "Service oriented architectures: approaches, technologies and research issues", The VLDB Journal, vol. 16, pp. 389-415.
[10] SAGA, Simple API for Grid Applications, http:/www.ogf.org/documents/GFD.90.pdf
[11] J. Pan, D. Rex, and A. Toga, "The LONI Pipeline Processing Environment: Improvements for Neuroimaging Analysis Research", 11th Annual Meeting of the Organization for Human Brain, 20051
[12] I. Altintas et al, "Kepler: an extensible system for design and execution of scientific workflows", Proceedings of the 16th International Conference on Scientific and Statistical Database Management, 2004, pp. 423- 424.